\documentclass[%
 reprint,
 amsmath,amssymb,
 aps,
 prl,
]{revtex4-2}

\usepackage{graphicx}
\usepackage{dcolumn}
\usepackage{bm}
\usepackage{multirow}
\usepackage{xcolor}
\usepackage[
    colorlinks=true,
    allcolors=blue,
    anchorcolor=blue,
    citecolor=blue,
    filecolor=blue,
    linkcolor=blue,
    urlcolor=blue
]{hyperref}
\usepackage{url}

\begin{document}

\preprint{APS/123-QED}

\title{The second altermagnet candidate in organic conductors: $\kappa$-(BEDT-TTF)$_2$$m$-HOOCC$_6$H$_4$SO$_3$.}

\author{Kohei Tokura$^{1,2}$}
\author{Takato Masuta$^1$}
\author{Kazushi Aoyama$^{2,3}$}
\author{Hiroki Akutsu$^{1,*}$}
\author{Yasuhiro Nakazawa$^1$}
\author{Scott S. Turner$^4$}

\affiliation{$^1$Department of Chemistry, The University of Osaka, Osaka 560-0043, Japan}
\affiliation{$^2$Department of Earth and Space Science, The University of Osaka, Osaka 560-0043, Japan}
\affiliation{$^3$Graduate School of Advanced Science and Engineering, Hiroshima University, Higashihiroshima 739-8521, Japan}
\affiliation{$^4$School of Chemistry and Chemical Engineering, University of Surrey, Guildford GU2 7XH, U.K.}

\date{\today}

\begin{abstract}
We have developed a novel BEDT-TTF-based organic conductor, $\kappa$-(BEDT-TTF)$_2$$m$-HOOCC$_6$H$_4$SO$3$ ($\kappa$-$m$-SBA), and propose it as a candidate for altermagnet.
Tight-binding band calculations of $\kappa$-$m$-SBA provide a $t'/t$ of 1.01 at 100 K, indicating that the spin structure is closely aligned to an equilateral triangle ($t'/t= 1$).
While most $\kappa$-type BEDT-TTF-based salts become spin liquids due to the spin frustration caused by the triangular lattice, $\kappa$-$m$-SBA surprisingly shows a weak ferromagnetic transition at $T_\text{N} = 14$ K due to a canted antiferromagnetic (AFM) spin structure. 
Until recently, $\kappa$-(BEDT-TTF)$_2$Cu[N(CN)$_2$]Cl ($\kappa$-Cl) was the only $\kappa$-type organic conductor known to exhibit this order, and it is also recognized as the first candidate for altermagnetism in organic conductors.
This was theoretically predicted by Naka $et  \ al$. in 2019, who demonstrated that $\kappa$-type organic conductors can be candidates for altermagnetism if they display such order. 
Consequently, $\kappa$-$m$-SBA can be considered the second candidate for altermagnetism in organic conductors.
Furthermore, numerical calculations demonstrate a characteristic of altermagnets in $\kappa$-$m$-SBA, namely spin splitting of energy bands.
\end{abstract}

\maketitle

A new class of magnetism, altermagnetism, has recently attracted great interest for both its fundamental physics and its potential utility in spintronics \cite{Smejkal2022-1,Smejkal2022-2}.
Altermagnets are antiferromagnets that break macroscopic time-reversal symmetry because their up- and down-spin sub-lattices are related by symmetry operations such as glide mirrors or screw rotations, rather than by translation or inversion.
This symmetry breaking induces a non-relativistic spin splitting of energy bands, originating from the cooperative effect of the magnetic order and sublattice-dependent anisotropic electron hopping.
Such features enable efficient spin current generation without net magnetization, making altermagnets ideal for high-speed, low-energy memory devices.

\begin{figure}[tbp]
    \centering
    \includegraphics[width=0.25\textwidth]{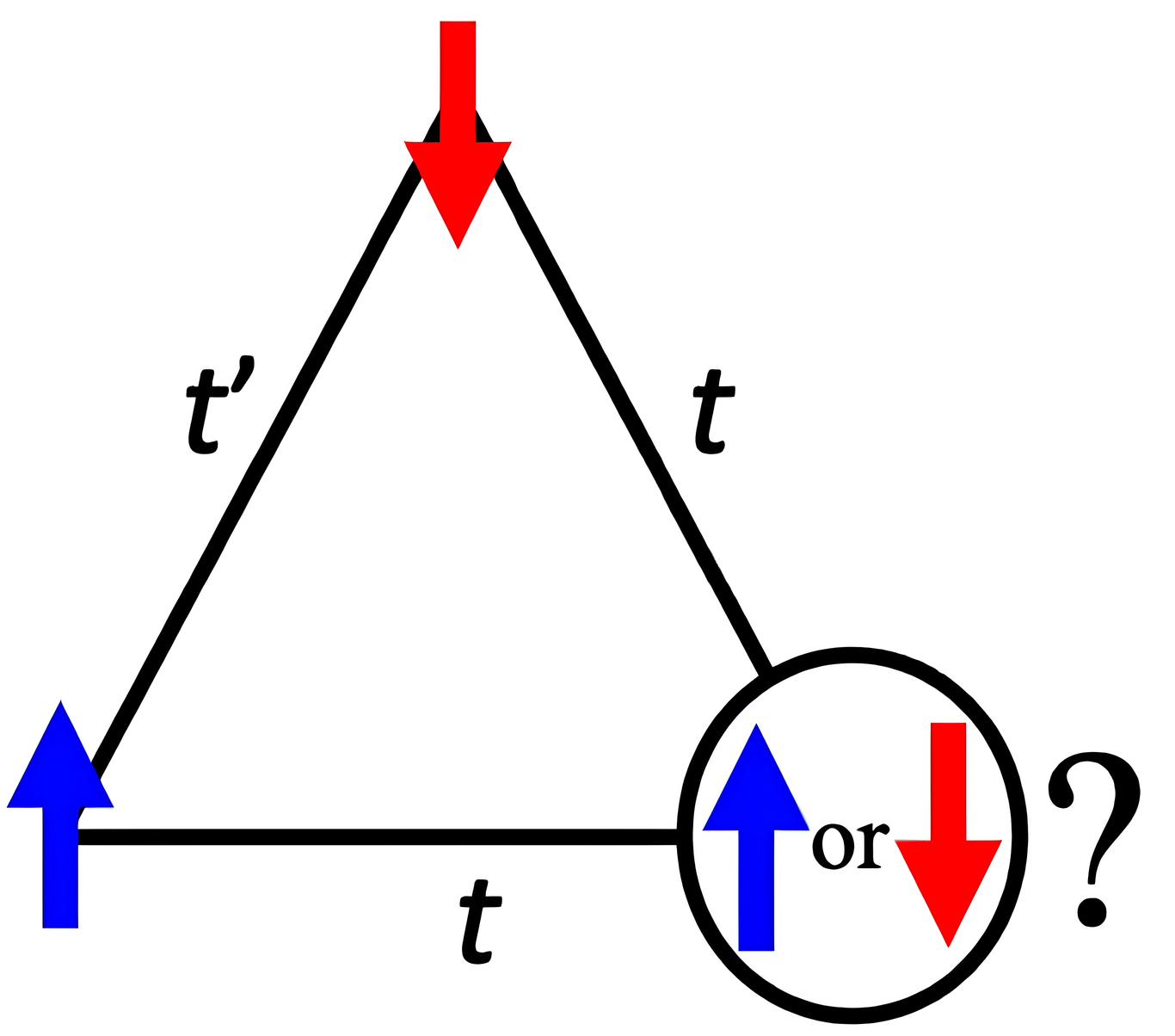}
    \caption{Frustration of spins on a triangular lattice.}
    \label{fig:frustration}
\end{figure}

While most currently identified altermagnets are inorganic materials, organic conductors are particularly promising as they were among the first systems where such altermagnetic effects were theoretically proposed.
Until now, the sole candidate for an altermagnetic organic conductor has been $\kappa$-(BEDT-TTF)$_2$Cu[N(CN)$_2$]Cl ($\kappa$-Cl), where BEDT-TTF = bis(ethylenedithio)tetrathiafulvalene. 
This was theoretically predicted by Naka et al. \cite{Naka2019}, who demonstrated that $\kappa$-type organic conductors generally manifest altermagnetism provided they exhibit collinear AFM. 
However, most $\kappa$-type salts do not show an AFM transition because of frustration due to a triangular lattice prescribed by the spins of the donor layer (Figure \ref{fig:frustration}) \cite{Pustogow2022}. 
As shown in Figure \ref{fig:frustration}, if two spins form a spin dimer at the corners of an equilateral triangular lattice ($t'/t = 1$), the third spin cannot participate in forming a spin dimer. 
This frustration makes most $\kappa$-type salts unable to exhibit magnetic order; and ultimately they become spin liquids without a static spin pattern,despite the spin-spin interactions being large. 
On the other hand, $\kappa$-Cl shows a canted AFM spin structure and a weak ferromagnetic transition at 22 K, which is attributed to a lattice distortions away from the perfect equilateral triangular geometry ($t'/t = 0.73$) \cite{Kawamoto2018}.

In this paper, we have developed the new organic conductor $\kappa$-(BEDT-TTF)$_2$$m$-HOOCC$_6$H$_4$SO$3$ ($\kappa$-$m$-SBA) and found that it exhibits a canted AFM spin structure below 14 K, similar to that of $\kappa$-Cl, although its $t'/t$ is 1.01 at 100 K, which almost describes a perfect equilateral triangular lattice.
Here we report the structure, properties of the $\kappa$-$m$-SBA salt, and discuss its candidacy as an altermagnet, including numerical calculations of the spin-split band structure.

\begin{figure}[tbp]
    \centering
    \includegraphics[width=\columnwidth]{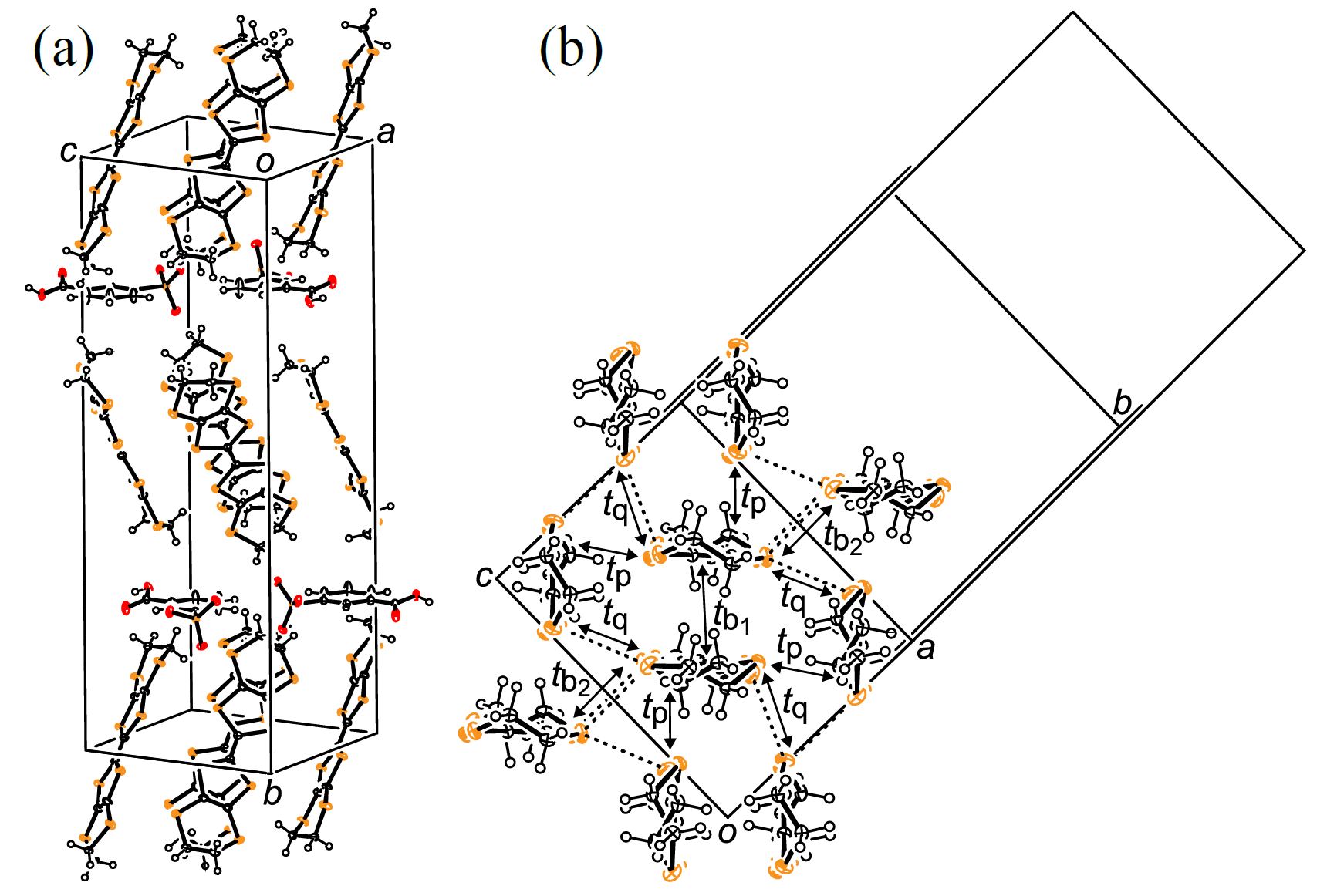}
    \caption{(a) Crystal structure of $\kappa$-$m$-SBA, where disorders are omitted for clarity. (b) Structure of the $\kappa$-type donor arrangement.}
    \label{fig:structure}
\end{figure}

Black rhombic $\kappa$-$m$-SBA crystals were prepared by a conventional electrocrystallisation method in a mixture of 16~mL of PhCl and 4~mL of EtOH with 15~mg of BEDT-TTF, 70~mg of $m$-HOOCC$_6$H$_4$SO$_3$Na (Tokyo Chemical Industry) and 100~mg of 18-crown-6 ether. 
The crystal structure is shown in Figure \ref{fig:structure}a. 
The structure of the $\kappa$-type donor arrangement is shown in Figure \ref{fig:structure}b. 
The $\kappa$-$m$-SBA salt is isostructural to $\kappa$-Cl, although the cell volume of $\kappa$-$m$-SBA is approximately 400~\AA$^3$ (10\%) larger than that of $\kappa$-Cl (Table S1) \cite{Geiser1991}. 
Both salts crystalize in the $P_{nma}$ space group and have one donor and a half of an anion in the asymmetric unit. 
For $\kappa$-$m$-SBA, the $c$ axis is only 0.5~\AA\ longer than the $a$ axis, unlike $\kappa$-Cl for which the $c$ axis is 4.5~\AA\ shorter than the $a$ axis. 
In addition, for $\kappa$-$m$-SBA the donor and the -COOH and -SO$_3^-$ groups of the anion have conformational disorders, as shown in Figure \ref{fig:S1}.
Tight-binding band calculations were performed for $\kappa$-$m$-SBA \cite{Mori1984}. 
\begin{table}[b]
\caption{Calculated band parameters.}
\label{tab:band_parameters}
\begin{ruledtabular}
\begin{tabular}{lcc}
Salt & $\kappa$-Cl$^{a)}$ & $\kappa$-$m$-SBA$^{c)}$ \\
\hline
$t_{\text{b1}}/10^{-3}$ eV & 27.3$^{b)}$ & 23.4 \\
$t_{\text{b2}}/10^{-3}$ eV & 10.4$^{b)}$ & 9.7 \\
$t_{\text{p}}/10^{-3}$ eV    & 10.5$^{b)}$ & 5.0 \\
$t_{\text{q}}/10^{-3}$ eV    & $-3.9^{b)}$ & $-4.6$ \\
$U_{\text{eff}}/10^{-3}$ eV & 54.6 & 46.8 \\
$W_{\text{u}}/10^{-3}$ eV   & 59.0 & 29.7 \\
$t'/t$                  & 0.72 & 1.01 \\
$U_{\text{eff}}/W_{\text{u}}$    & 0.93 & 1.58 \\
$U_{\text{eff}}/t$      & 7.6  & 10.1 \\
\end{tabular}
\end{ruledtabular}
\footnotesize
$^{a)}$at 127~K \cite{Mori1999}. \\
$^{b, c)}$ at 100~K.
\end{table}
Due to the positional disorder of an outer sulfur atom over two positions (Figure \ref{fig:S1}), two sets of transfer integrals were calculated (Table S2), the average of which was used to calculate the band dispersions and Fermi surfaces, which are shown in Figure \ref{fig:S3}. 
A wide Mott gap of 0.22~eV is observed, which is similar to that of $\kappa$-Cl (0.25~eV), whose band dispersions and Fermi surfaces are shown in Figure \ref{fig:S4}. 
The Fermi surfaces are quite similar to those of $\kappa$-Cl and other $\kappa$-type salts. 
The two upper bands (upper two dispersion curves) are also quite similar, but the two lower bands are quite different. 
Table \ref{tab:band_parameters} shows calculated band parameters using the Tight-binding method.
The $t_p$ and $W_{\mathrm{u}}$ of $\kappa$-$m$-SBA are approximately half of those of $\kappa$-Cl, indicating that $\kappa$-$m$-SBA is a poorer conductor than $\kappa$-Cl.
Moreover, the frustration parameter, $t'/t$ for $\kappa$-$m$-SBA is 1.01, which is the average calculated from the two disordered structures (A: 0.970; B: 1.055). 
This value indicates that the triangular lattice is very close to being equilateral, which would typically reslut in $\kappa$-$m$-SBA being a spin liquid. 
Among the four transfer integrals, $t_{b1}$ is the largest, suggesting that each spin is localized on the BEDT-TTF face-to-face dimer. 
In addition, a first-principles band calculation was performed using OpenMX \cite{Weng2009}, which provides a $t'/t$ of 0.80.

The temperature dependence of electrical resistivity was measured by a conventional four-probe method. 
$\kappa$-$m$-SBA is a semiconductor with a room temperature resistivity ($\rho_{\text{RT}}$) of $280\,\Omega\cdot\text{cm}$ and activation energy ($E_a$) of $43\,\text{meV}$. 
In comparison, the $\rho_{\text{RT}}$ ($\sim 0.5\,\Omega\cdot\text{cm}$) and $E_a$ ($12\,\text{meV}$) of $\kappa$-Cl \cite{Williams1990} are approximately 500 and 4 times lower than those of $\kappa$-$m$-SBA, respectively. 
This experimentally confirms that $\kappa$-$m$-SBA is poorer conductor than $\kappa$-Cl.

\begin{figure}[b]
    \centering
    \includegraphics[width=0.45\textwidth]{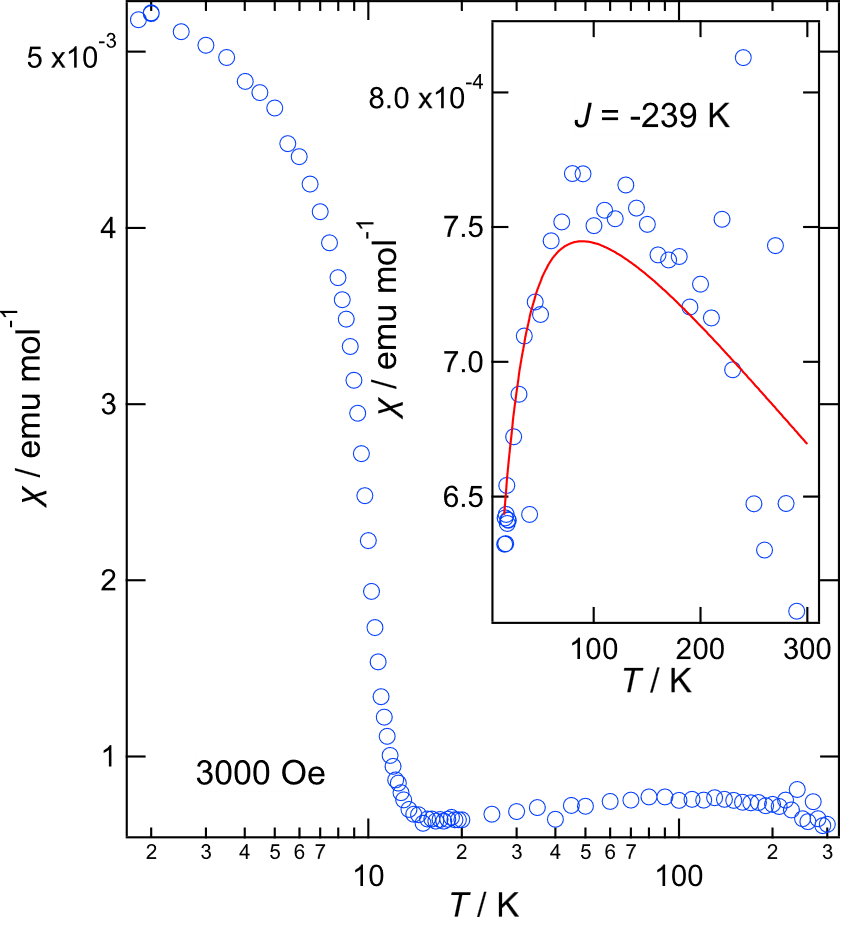}
    \caption{$\chi$-$T$ plots of $\kappa$-$m$-SBA. The inset shows the expansion from 16-300\,K.}
    \label{fig:susceptibility}
\end{figure}
\begin{figure}[b]
    \centering
    \includegraphics[width=0.45\textwidth]{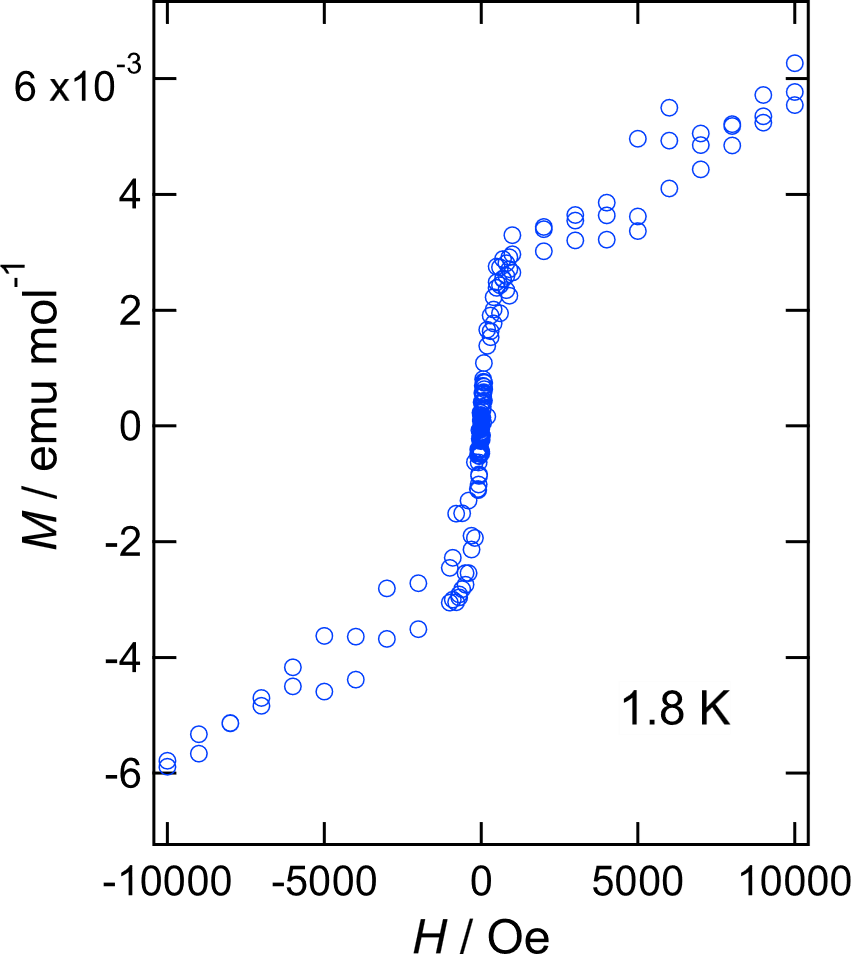}
    \caption{$M$-$H$ plots of $\kappa$-$m$-SBA.}
    \label{fig:magnetization}
\end{figure}

Temperature-dependent and field-dependent magnetic susceptibilities were measured using a Quantum Design MPMS-XL SQUID magnetometer. 
Figure \ref{fig:susceptibility} shows the temperature dependence of magnetic susceptibility from 1.8-300\,K. 
An abrupt increase is observed at 14\,K, indicating a ferromagnetic transition. 
The inset of Figure \ref{fig:susceptibility} is an expansion from 16-300\,K. 
The plots can be fitted using a triangular lattice model with $J = -239$\,K. 
The behavior is quite similar to that of a typical spin liquid.
Figure \ref{fig:magnetization} shows the field-dependent-magnetization. 
A small saturation magnetization of 0.3\% is observed, which is larger than that of the isomorphous canted AFM $\kappa$-Cl (0.08\%) \cite{Welp1993} but still much lower than that of a normal ferromagnet. 
The material also appears to be a soft magnet since the hysteresis loop is very narrow as shown in Figure \ref{fig:magnetization} and Figure \ref{fig:S5}. 
These results definitely show that $\kappa$-$m$-SBA is a canted AFM similar to $\kappa$-Cl and is only the second canted AFM discovered in the $\kappa$-family.

Here we speculate why $\kappa$-mSBA, possessing an almost equilateral triangular lattice, does not become a spin liquid but is a canted AFM. 
To support these assertions, we have performed first-principles band calculations using the OpenMX software package with the Perdew-Burke-Erzerhof (PBE) type generalized gradient approximation (GGA).
Figure \ref{fig:S6} shows band dispersions, which are similar to that from the tight-binding (TB) results. 
Figure \ref{fig:S7} shows the calculated Fermi surfaces, the areas of which are smaller than those using the TB method.
All transfer integrals are shown in Table S2. 
A $t'/t$ of 0.80 was calculated, which is smaller than that from the TB method and smaller than 1.00, indicating the deviation from a perfect equilateral triangle. 
So, this seems to be the reason why this material is a bulk weak ferromagnet resulting from the canted AFM spin geometry.
However, a $t'/t \approx 0.80$ seems to be commonly seen for first principles (FP) studies of spin liquids. 
For example, a report\cite{Kandpal2009} of the FP study stated that $t'/t$ values of TB and FP are 0.75 and 0.42 for $\kappa$-Cl, respectively, but 1.06 and 0.83 for a typical spin liquid such as $\kappa$-CN ($\kappa$-(BEDT-TTF)$_2$Cu$_2$(CN)$_3$). 
The $t'/t$ value using FP of the typical spin liquid of 0.83 is quite close to that calculated for $\kappa$-mSBA of 0.80.
We also calculated the FP band structures using only donor molecules with a uniform background charge of $-4$. 
For the additional calculations, we estimated the effect of anions on $t'/t$. 
The values are also listed in Table S2. 
The $t'/t$ value using FP with only donors of 0.86, which is 0.06 larger than that of the same calculation using the whole crystal (including anions). 
This suggests that the presence of the anions makes the triangular lattice diverge from being equilateral.
The anion, HOOCC$_6$H$_4$SO$_3^-$, is large but the negative charge is localized on a specific part, the -SO$_3^-$ group, and the isolated anion has a relatively large dipole moment of 12.8~debye\cite{Stewart2016} for both disorder structures. 
As shown in Figure \ref{fig:S8}, there are two anions having different orientations in the repeating unit. 
The dipole moments in both molecules are partially cancelled in the repeating unit and 11.1~debye of a dipole moment per anion remains along the $a$-axis.
Therefore, the localization of the minus charge on the specific functional group together with the dipole moment of the anionic layers polarized along the $a$-axis provide a steep electric field (Figure \ref{fig:S8}).
This may change the shape of the electron clouds on the BEDT-TTF molecules. 
The resultant deformed electron clouds might then provide a distorted triangular lattice, which consequently allows the magnetic order.

In light of the experimental results and the first-principles findings described above, we now discuss the possibility of altermagnetism in $\kappa$-$m$-SBA. 
Theoretical studies\cite{Naka2019} have demonstrated that the $\kappa$-type arrangement serves as a platform for altermagnetism, provided the system exhibits collinear AFM order.
Although the magnetic structure of $\kappa$-$m$-SBA involves a slight canting as discussed above, it is theoretically effective to approximate the system as a collinear AFM to evaluate its potential for altermagnetism, which is primarily governed by the lattice symmetry.
On this basis, $\kappa$-$m$-SBA satisfies the essential requirements for an altermagnet candidate.
To further elucidate the electronic structure resulting from this magnetic state, numerical calculations were done to investigate the spin-dependent band structure under the collinear AFM order\cite{Noda2016,Okugawa2018,Hayami2018,Yuan2020}.

\begin{figure}[b]
    \centering
    \includegraphics[width=0.5\textwidth]{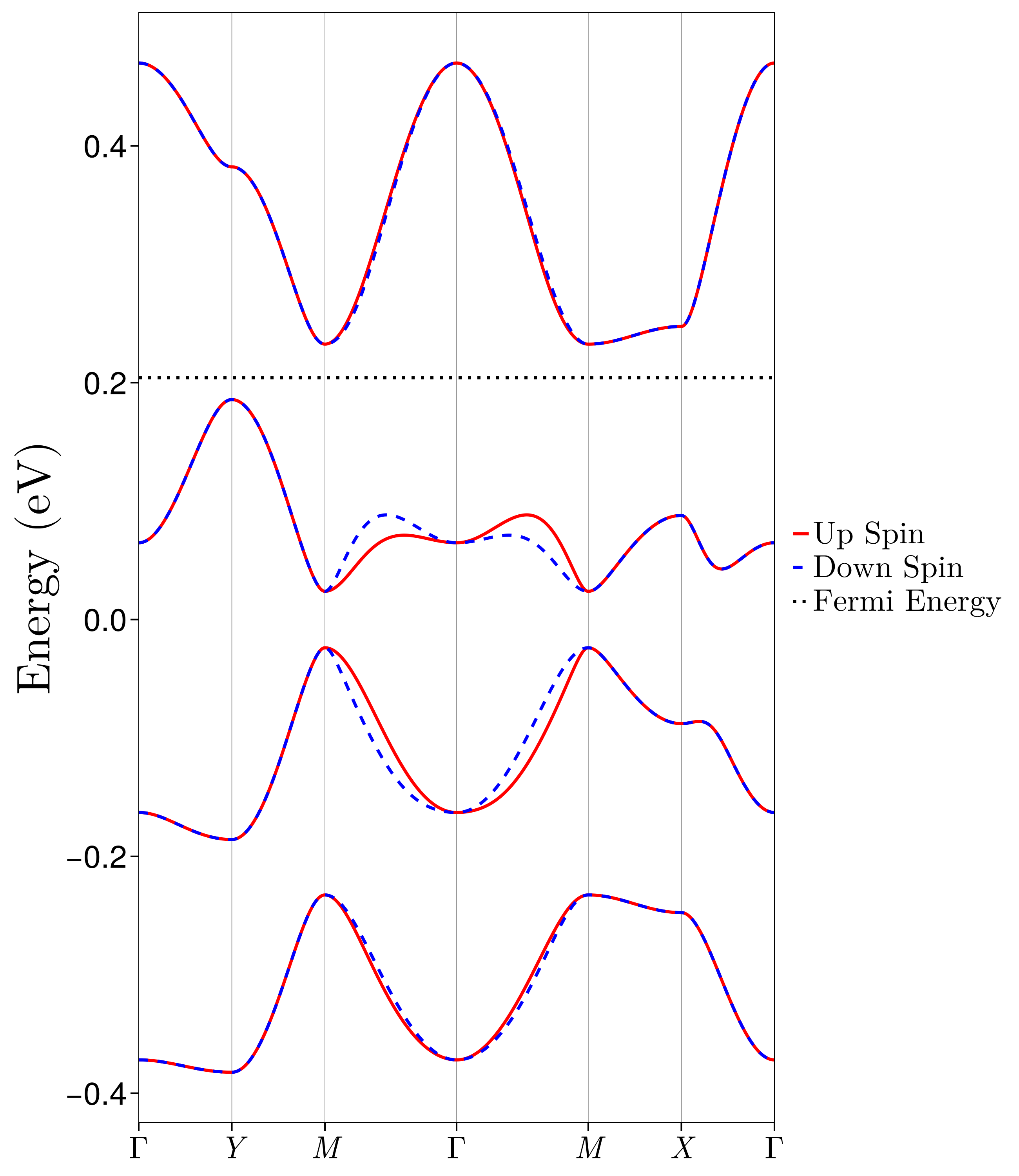}
    \caption{The energy bands of $\kappa$-$m$-SBA. }
    \label{fig:bands}
\end{figure}

We investigate the spin-dependent band structure of $\kappa$-$m$-SBA based on the Hubbard model, taking into account the two distinct types of dimers and the anisotropy in the transfer integrals between them, $\left(t_{\mathrm{b1}},t_{\mathrm{b2}},t_{\mathrm{p}},t_{\mathrm{q}}\right) = \left(198, 70, 44, -42\right)$ meV, based on first-principles calculations, where the opposite sign of $t_{\mathrm{p}}$ is chosen according to the phase convention adopted in the present effective model\cite{Seo2004,Koretsune2014}.
The Hamiltonian is given by
\begin{widetext}
\begin{equation}
    \mathcal{H}_{\text{Hubb}} = 
    t_{b1} \sum_{i\sigma} \left( \hat{c}_{ia\sigma}^\dagger \hat{c}_{ib\sigma} + \text{h.c.} \right) 
    + \sum_{\substack{\langle i, j \rangle \\ \mu, \mu', \sigma} } t_{ij}^{\mu\mu'} \left( \hat{c}_{i\mu\sigma}^\dagger \hat{c}_{j\mu'\sigma} + \text{h.c.} \right) 
    + U \sum_{i, \mu} \hat{n}_{i\mu\uparrow} \hat{n}_{i\mu\downarrow} 
\end{equation}
\end{widetext}
where $t_{b1}$ is the intra-dimer transfer integral, and $t_{ij}^{\mu\mu'}$ denotes the inter-dimer transfer integrals ($t_{ij}^{\mu\mu'} = \left\{ t_{b2}, t_{p}, t_{q} \right\} $).
The last term in Eq.~(1) represents the intra-molecular Coulomb interaction. 
While Ref.~\cite{Naka2019} adopted $U=1$ eV, here we use $U=0.54$ eV to reflect the reduced electronic energy scale of $\kappa$-$m$-SBA: $t_p$ and $W_{\mathrm{u}}$ are both approximately half of those of $\kappa$-Cl.
By using the mean-field approximation $n_{i\mu\uparrow}n_{i\mu\downarrow} \simeq n_{i\mu\uparrow}\langle n_{i\mu\downarrow}\rangle + \langle n_{i\mu\uparrow}\rangle n_{i\mu\downarrow} - \langle n_{i\mu\uparrow}\rangle \langle n_{i\mu\downarrow}\rangle$, 
we calculate the energy dispersion for the 3/4 filling, where $\langle n_{i\mu\uparrow}\rangle$ and $\langle n_{i\mu\downarrow}\rangle$ are determined self-consistently within two-sublattice AFM configurations.
Figure \ref{fig:bands} shows the energy band dispersion in the AFM insulating phase, where, as in Ref.~\cite{Naka2019}, the lattice is simplified to be a square lattice and $\delta=\langle n_{\uparrow}\rangle-\langle n_{\downarrow}\rangle$ turns out to be nonzero, signaling the occurrence of AFM order.
Consistent with the results reported for $\kappa$-Cl (cf. Fig.2d in Ref.\cite{Naka2019}), a clear spin splitting is observed along the path connecting the $\Gamma$ and M points. 
This splitting originates from the cooperative effect of the AFM ordering and the sublattice-dependent anisotropic electron hoppings.
This confirms that $\kappa$-$m$-SBA manifests the characteristic electronic signature of an altermagnet.

In addition, for more than 15 years, we have developed polar organic conductors, in which each polar counterion aligns in the same orientation as the other ions in the same layer. 
These salts are classified into four types as shown in Figure \ref{fig:S9}. 
$\kappa$-$m$-SBA belongs to Type III \cite{Akutsu2022}, which possesses no net dipole moment for the bulk crystal, but each anionic layer has a finite dipole moment. 
This non-centrosymmetric nature within the layers is likely to support the presence of the Dzyaloshinskii--Moriya interaction, leading to the observed weak ferromagnetism.

In summary, $\kappa$-$m$-SBA is the second canted AFM in the $\kappa$-type organic conductors with $T_N = 14$\,K. 
Consequently, it represents the second candidate for an altermagnet in this family of materials.
It is thus a first purely organic altermagnet candidate.

We are grateful to Makoto Naka for the fruitful discussion.
H.A. would like to thank the Izumi foundation (2023-J-077) for financial support in part. 
T.M. appreciates financial support by JST SPRING, Japan Grant (Number JPMJSP2138).
K.A. is supported by JSPS KAKENHI Grant No. JP23H00257 and JP24K00572.

\bibliography{references}

\clearpage
\appendix
\onecolumngrid

\setcounter{table}{0}
\setcounter{figure}{0}
\renewcommand{\thetable}{S\arabic{table}}
\renewcommand{\thefigure}{S\arabic{figure}}
\renewcommand{\topfraction}{1.0}
\renewcommand{\bottomfraction}{1.0}
\renewcommand{\textfraction}{0.0}
\renewcommand{\floatpagefraction}{1.0}

\section*{Supporting Information}

\begin{table}[htbp]
\centering
\caption{X-ray crystallographic data of $\kappa$-mSBA at 100 and 293~K together with $\kappa$-Cl at 127~K \cite{Geiser1991}.}
\label{tab:S1}
\begin{ruledtabular}
\begin{tabular}{lccc}
 & \multicolumn{2}{c}{$\kappa$-mSBA} & $\kappa$-Cl \\
\hline
Composition & \multicolumn{2}{c}{C$_{27}$H$_{21}$O$_5$S$_{17}$} & C$_{22}$H$_{19}$S$_{16}$ClCu \\
$T$ / K & 293 & 100 & 127 \\
Space group & $P_{nma}$ & $P_{nma}$ & $P_{nma}$ \\
$a$ / \AA & 10.6268(3) & 10.5421(4) & 12.909(3) \\
$b$ / \AA & 31.0785(9) & 30.9105(10) & 29.638(3) \\
$c$ / \AA & 11.1784(4) & 11.0372(4) & 8.418(1) \\
$V$ / \AA$^3$ & 3691.8(2) & 3596.6(2) & 3222.9(9) \\
$Z$ & 2 & 2 & 2 \\
$R$ & 0.044 & 0.035 & 0.050 \\
$wR$ & 0.115 & 0.085 & 0.043 \\
\end{tabular}
\end{ruledtabular}
\end{table}
\begin{table}[htbp]
\centering
\caption{Calculated band parameters of two disordered states of $\kappa$-mSBA at 100~K.}
\label{tab:S2}
\begin{ruledtabular}
\begin{tabular}{ccccccc} 
Method & Disorder pattern & $t_{\text{b1}}$ / meV & $t_{\text{b2}}$ / meV & $t_{\text{p}}$ / meV & $t_{\text{q}}$ / meV & $t'/t$ \\
\hline
\multirow{3}{*}{Tight binding\cite{Mori1999}} & Disorder A & 227 & 96 & 50 & $-50$ & 0.97 \\
 & Disorder B & 240 & 97 & 50 & $-43$ & 1.06 \\
 & Average & 234 & 97 & 50 & $-46$ & 1.01 \\
\hline
\multirow{3}{*}{\shortstack{First principles\cite{Ozaki2003} \\ BEDT-TTF only$^{*1}$}} & Disorder A & 189 & 71 & $-41$ & $-44$ & 0.97 \\
 & Disorder B & 192 & 72 & $-43$ & $-39$ & 1.06 \\
 & Average & 190 & 72 & $-42$ & $-42$ & 0.86 \\
\hline
\multirow{3}{*}{\shortstack{First principles\cite{Ozaki2003} \\ whole crystal}} & Disorder A & 197 & 67 & $-44$ & $-45$ & 0.76 \\
 & Disorder B & 199 & 72 & $-45$ & $-40$ & 0.85 \\
 & Average & 198 & 70 & $-44$ & $-42$ & 0.80 \\
\end{tabular}
\end{ruledtabular}
\begin{flushleft}
\footnotesize
$^{*1}$ The calculations were performed using only donors and mean-field charge.
\end{flushleft}
\end{table}

\begin{figure}[htbp]
    \centering
    \includegraphics[width=0.55\columnwidth]{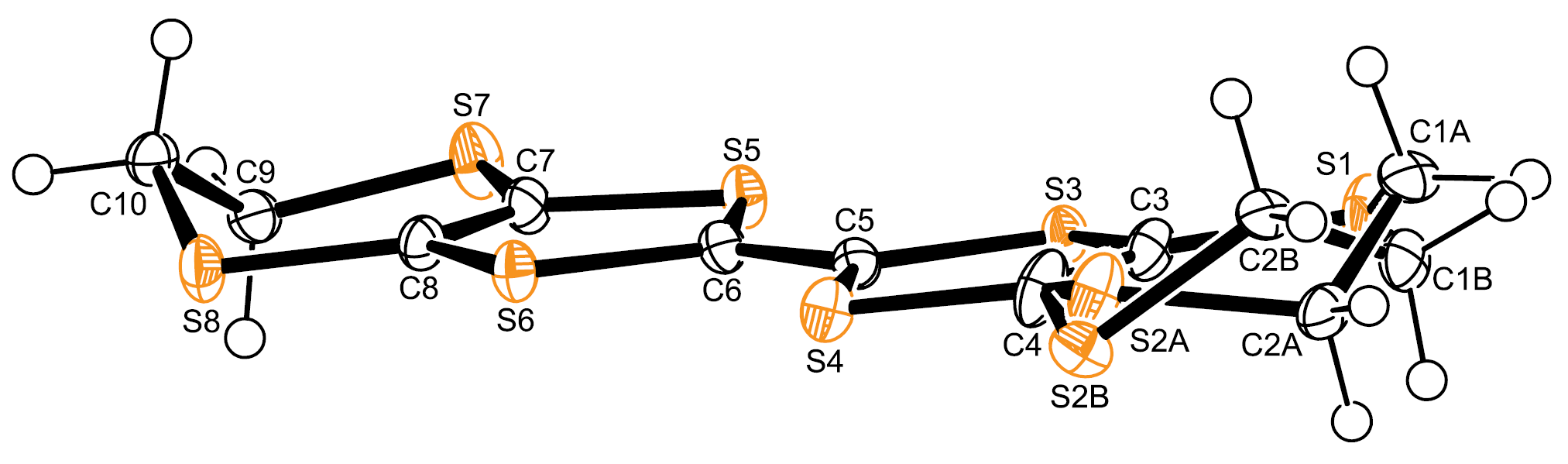}
    \caption{The molecular structure of BEDT-TTF.}
    \label{fig:S1}
\end{figure}
\begin{figure}[htbp]
    \centering
    \includegraphics[width=0.34\columnwidth]{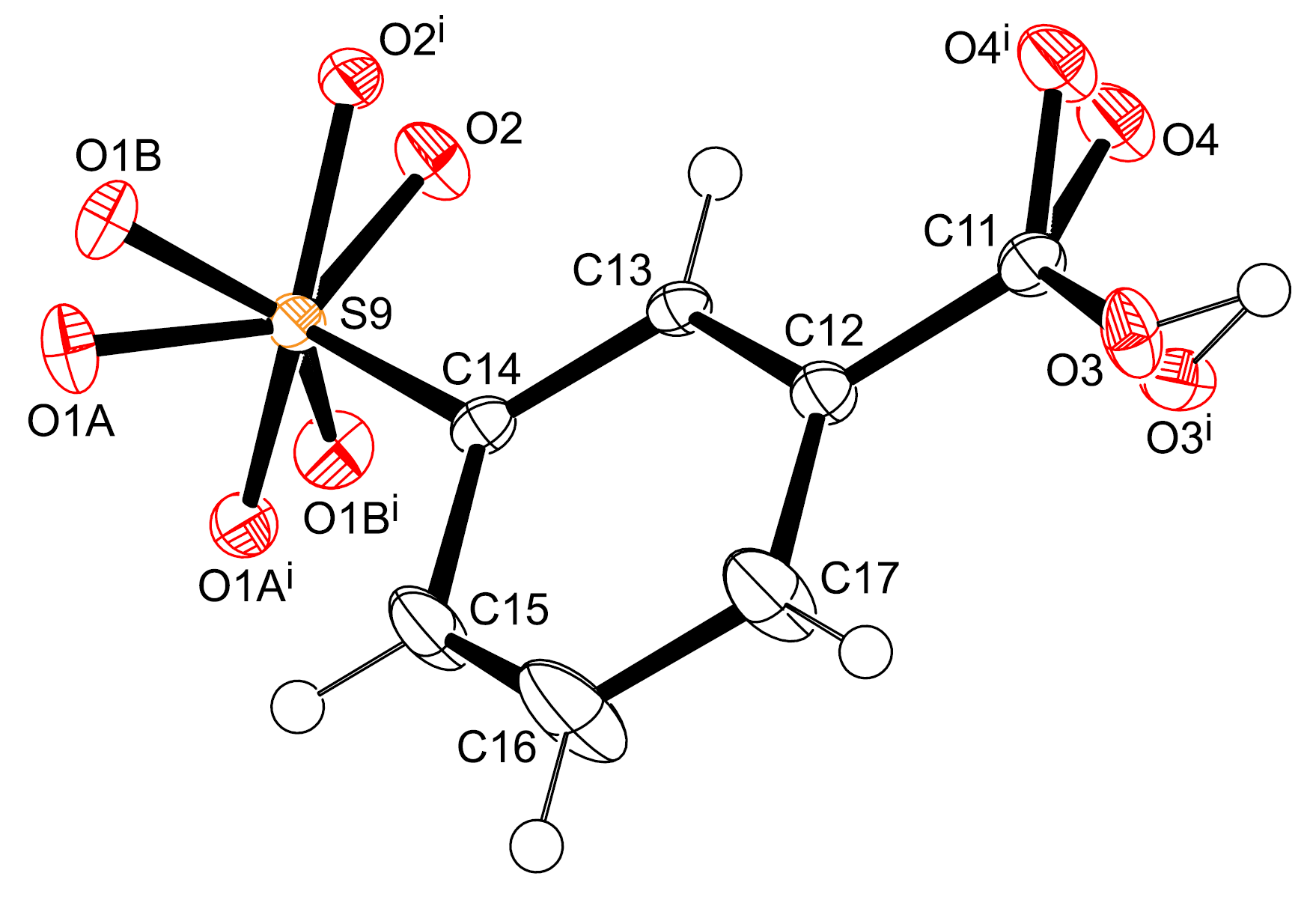}
    \caption{The molecular structure of the anion, which is located about a mirror plane. [Symmetry code: (i) $+x$, $1/2 - y$, $+z$]}
    \label{fig:S2}
\end{figure}
\begin{figure}[htbp]
    \centering
    \includegraphics[width=0.9\columnwidth]{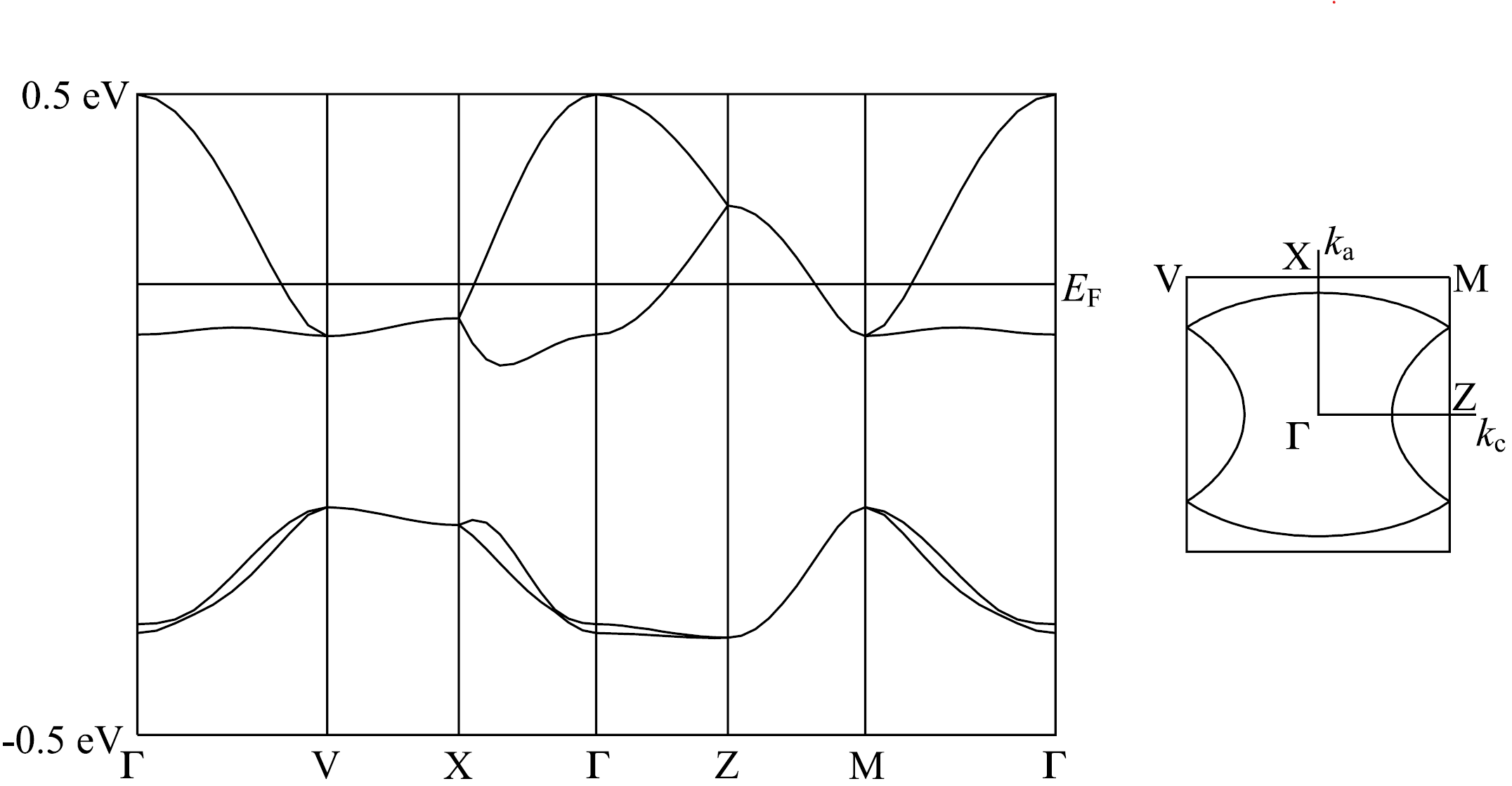}
    \caption{Band dispersions (left) and Fermi surfaces (right) of $\kappa$-$m$-SBA at 100\,K.}
    \label{fig:S3}
\end{figure}
\begin{figure}[htbp]
    \centering
    \includegraphics[width=0.9\columnwidth]{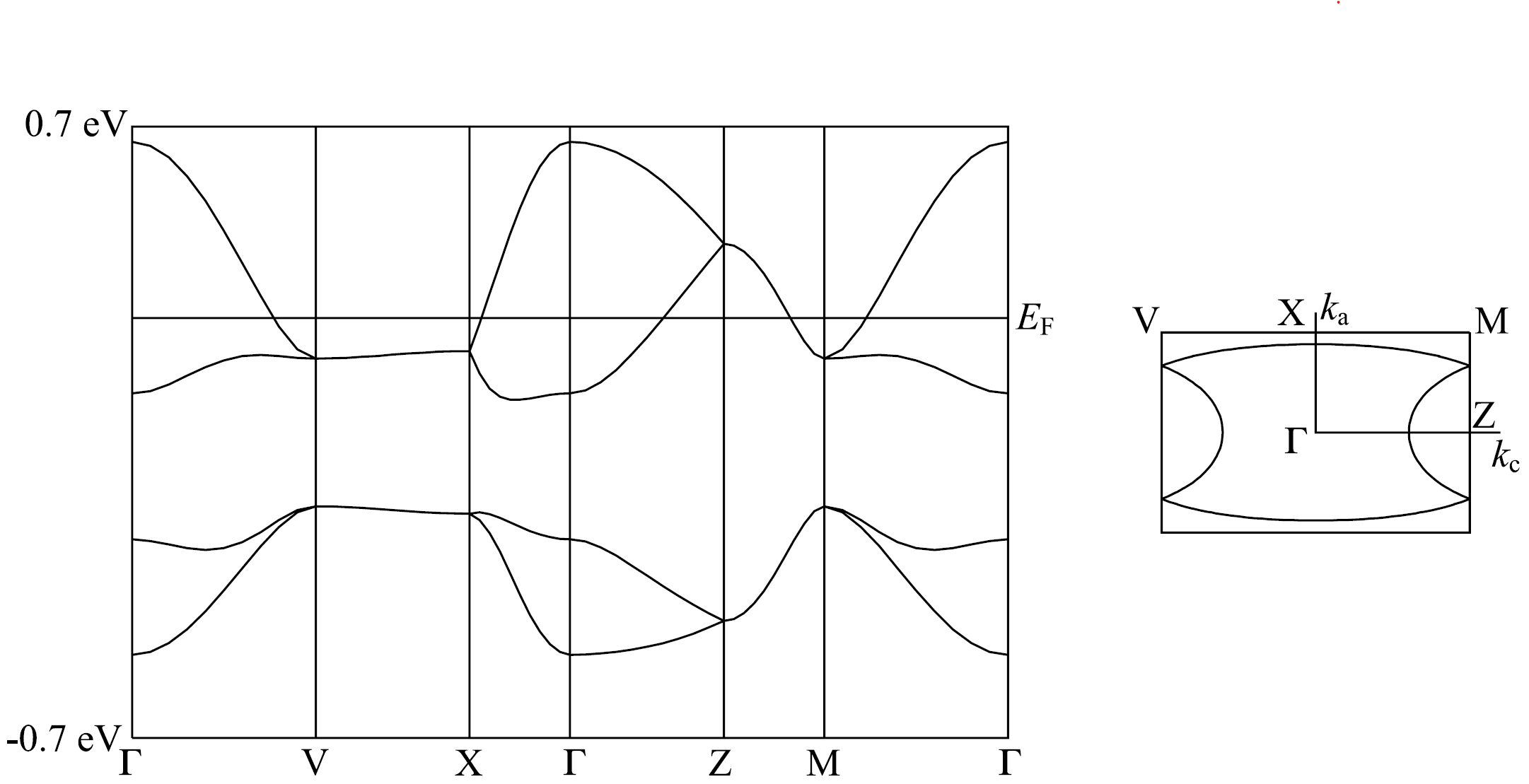}
    \caption{Band dispersions (left) and Fermi surfaces (right) of $\kappa$-Cl at 127\,K calculated using cell parameters and transfer integrals reported in ref. \cite{Mori1999}.}
    \label{fig:S4}
\end{figure}
\begin{figure}[htbp]
    \centering
    \includegraphics[width=0.6\columnwidth]{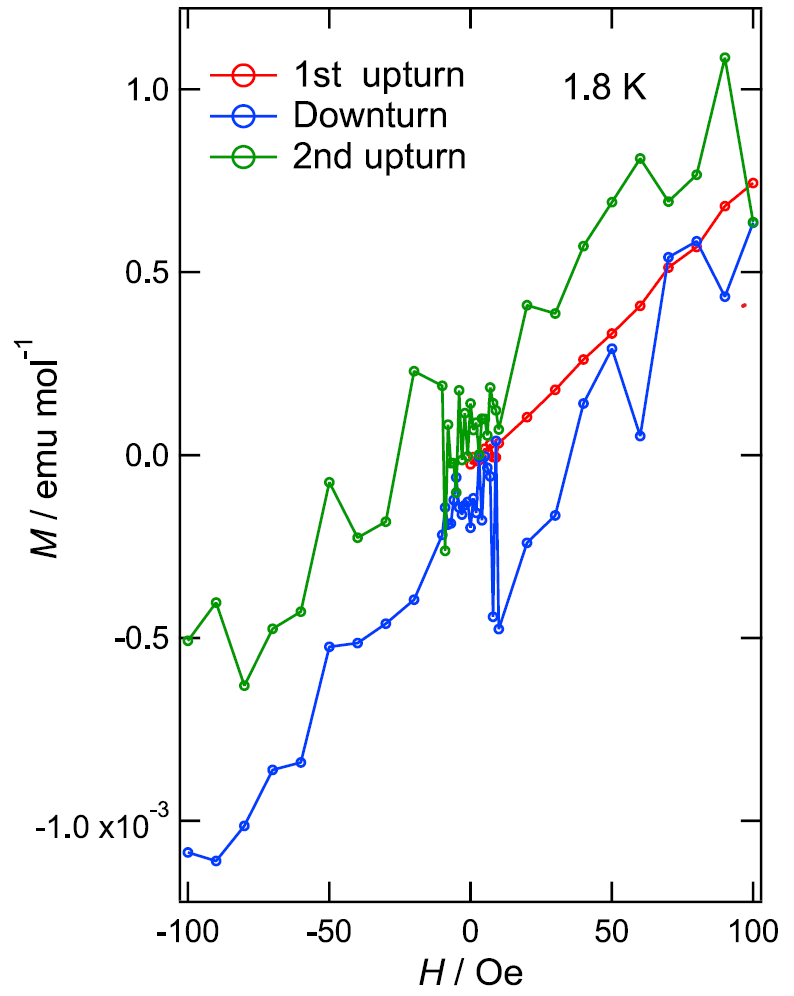}
    \caption{An expansion from $-100$ to $100$\,Oe of $M$--$H$ curves (Figure\ref{fig:susceptibility}) of $\kappa$-$m$-SBA at 1.8\,K.}
    \label{fig:S5}
\end{figure}
\begin{figure}[htbp]
    \centering
    \includegraphics[width=0.65\columnwidth]{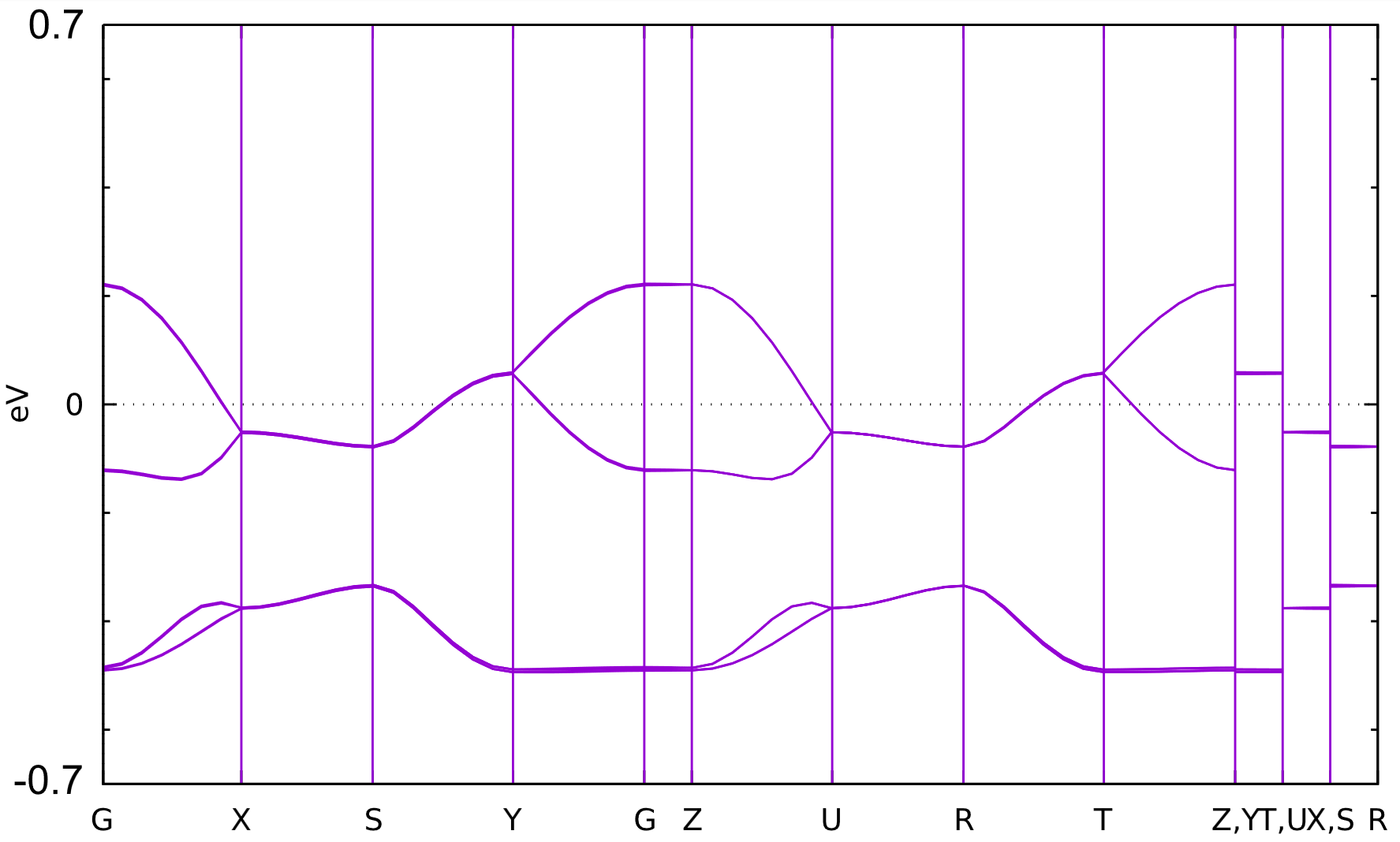}
    \caption{Band dispersion calculated by OpenMX of $\kappa$-mSBA (Disorder A) at 100 K.}
    \label{fig:S6}
\end{figure}
\begin{figure}[htbp]
    \centering
    \includegraphics[width=0.65\columnwidth]{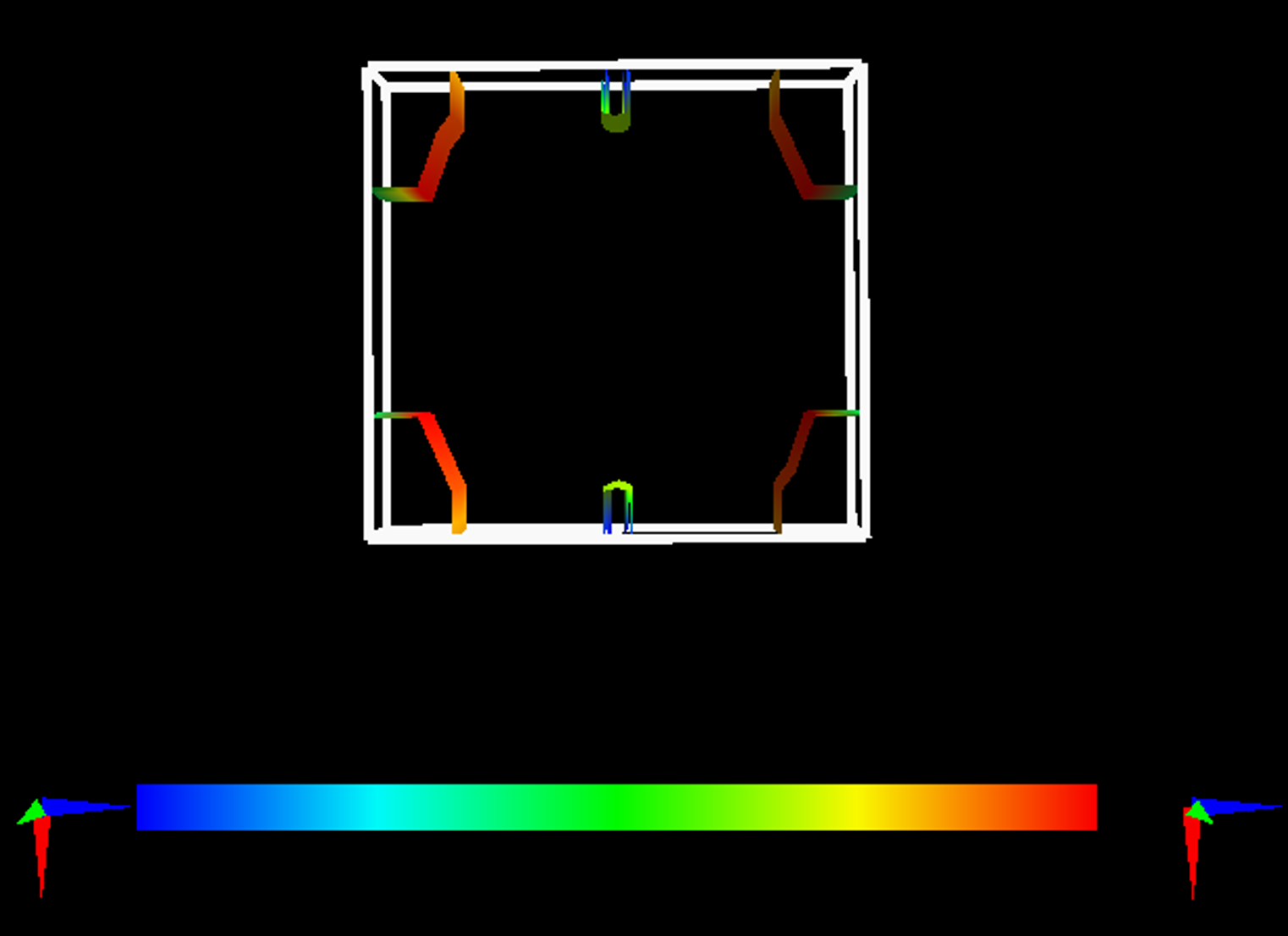}
    \caption{Fermi surfaces calculated by OpenMX of $\kappa$-mSBA (Disorder A) at 100 K.}
    \label{fig:S7}
\end{figure}
\begin{figure}[htbp]
    \centering
    \includegraphics[width=0.5\columnwidth]{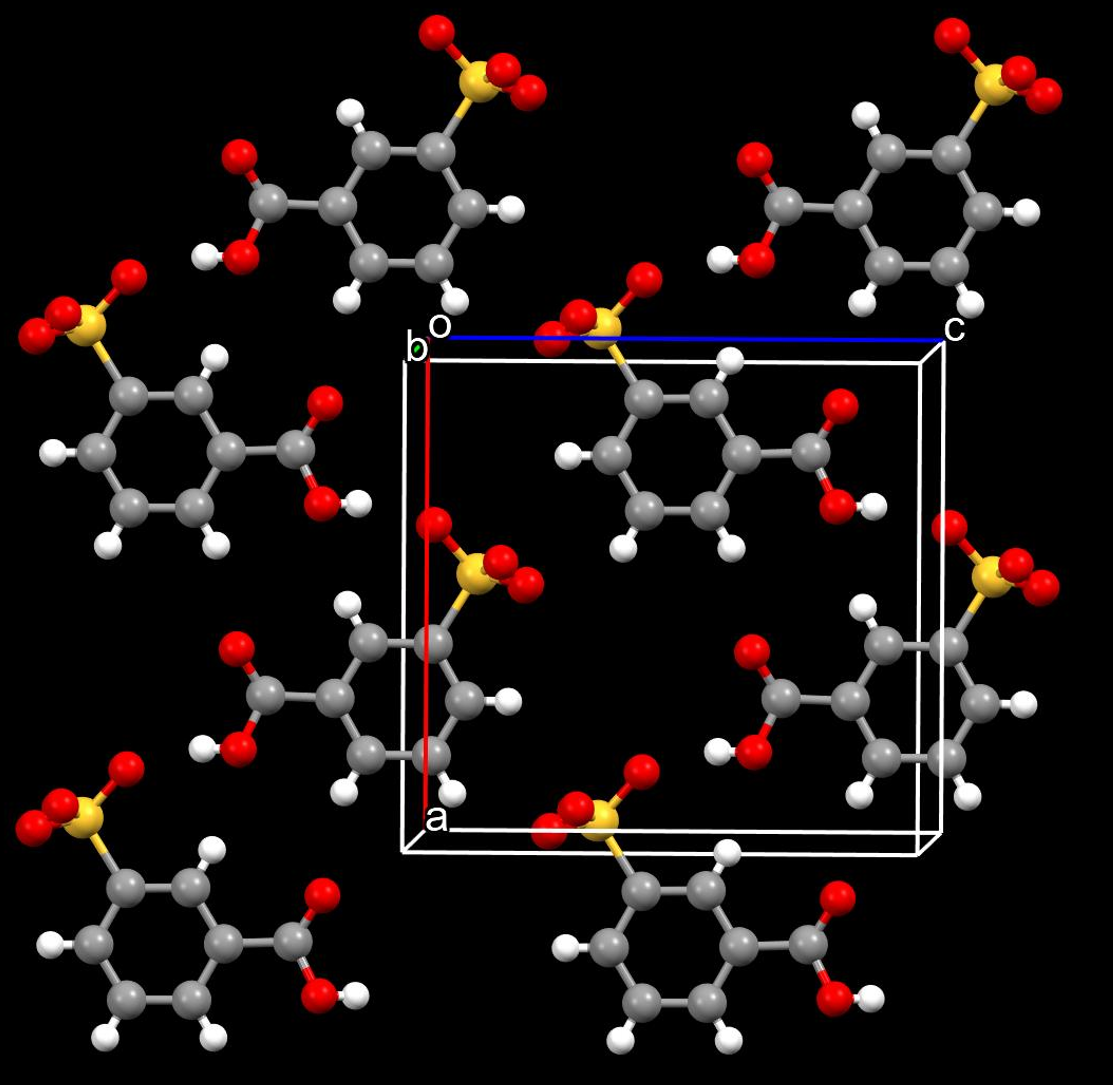}
    \caption{Crystal structure of an anionic layer of $\kappa$-mSBA (Disorder A) at 100~K.}
    \label{fig:S8}
\end{figure}
\begin{figure}[htbp]
    \centering
    \includegraphics[width=0.7\columnwidth]{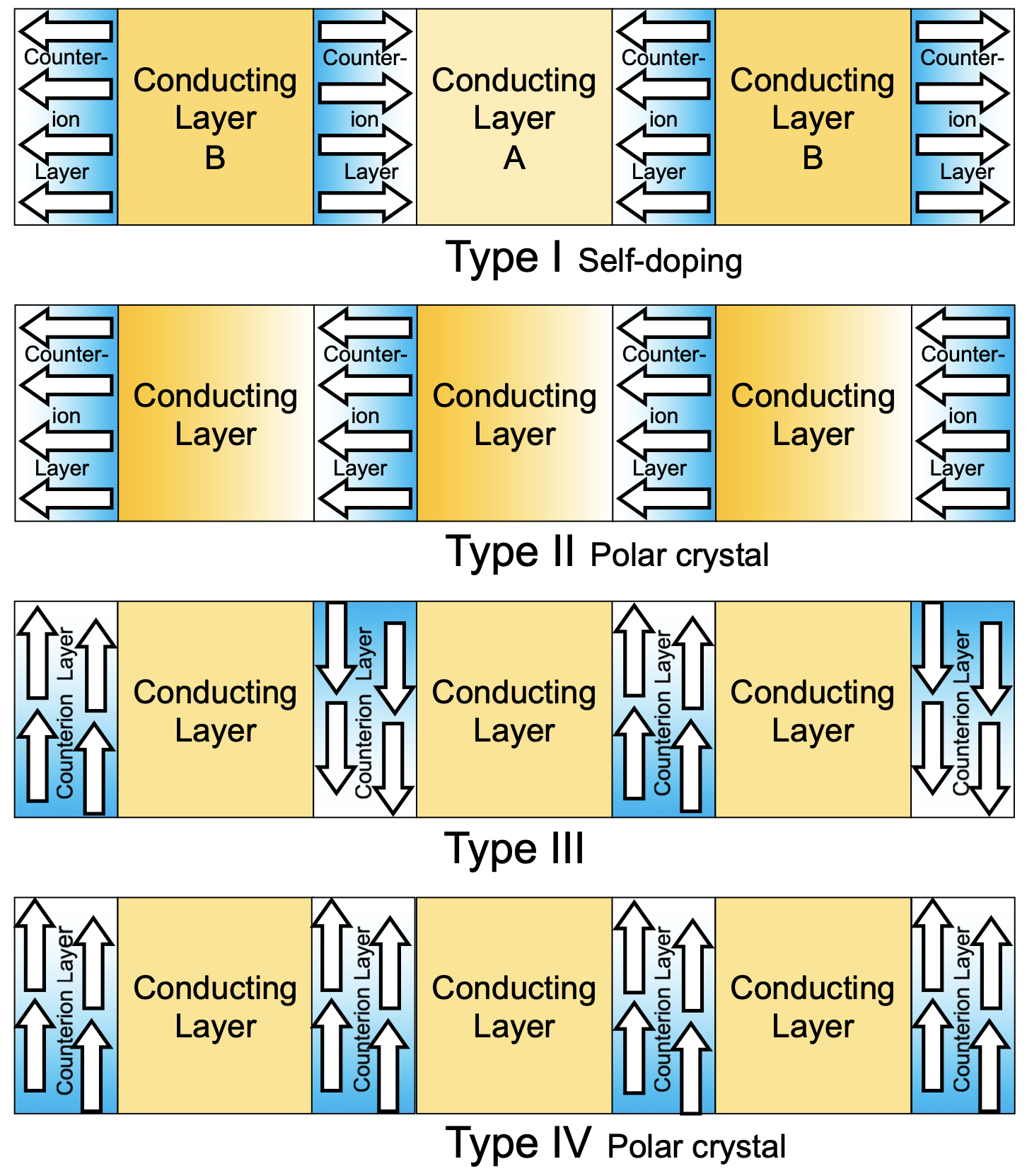}
    \caption{Schematic diagram of the crystal structures of polar organic conductors where the dipoles of the counterions are indicated by arrows (less negative $\Leftarrow$ more negative).}
    \label{fig:S9}
\end{figure}

\end{document}